\documentclass{aa}  
\usepackage{graphicx}
\usepackage{txfonts}
\usepackage{hyperref}
\hypersetup{
    colorlinks   = true,
    citecolor=blue,
    urlcolor=black
}% To add links in your PDF file, use the package "hyperref"
\usepackage{natbib}
\usepackage{amsmath}
\usepackage{booktabs}
\usepackage{multirow}
\usepackage{tablefootnote}
\usepackage{subfigure}
\usepackage{xcolor}

\definecolor{PineGreen}{RGB}{1, 121, 111}
\definecolor{RoyalBlue}{RGB}{65,105,225}

\def\Hii{H\,{\sc ii}}

\def\kms{km\,s$^{-1}$}

\def\msun{M$_{\odot}$}

\def\srv{$\sigma_\mathrm{RV}$}

\def\micron{$\mu$m}

\def\ha{H\,{\sc i}}

\def\hea{He\,{\sc i}}
\def\heb{He\,{\sc ii}}

\def\xsh{X-shooter}
\def\srv{$\sigma_\mathrm{1D}$}
\def\fbin{$f_\mathrm{bin}$}

\def\pcutoff{$P_\mathrm{cutoff}$}
\def\srvM{$8.9\pm0.5$}
\def\srvMorig{$7.4\pm0.6$}

\def\efoldtime{$0.15^{+0.06}_{-0.04}$}

\def\pcutoffM{415}
\def\pcutoffMoSig{85}
\def\pcutoffMtSig{30}
\def\hardtime{0.85}
\def\hardrate{3.4}
\def\hardrateOsig{1.6}

\def\obsFbin{$0.27\pm0.11$}
\def\intrFbin{87}
\def\intrPcutoff{50}

\def\pcutoffMfeighty{780}
\def\pcutoffMfeightyOsig{188}
\def\pcutoffMfeightyTsig{73}
\def\hardtimefeighty{1}
\def\hardratefeighty{4.4}
\def\hardratefeightyOsig{2.1}

\begin{document}

   \title{The spectroscopic binary fraction of the young stellar cluster M17\thanks{Based on observations collected at the European Southern Observatory at Paranal, Chile (ESO programs 0103.D-0099).}}

   \author{M.C. Ram\'irez-Tannus
      \inst{1}
    \and
    A.R. Derkink
    \inst{2}
    \and
    F. Backs 
    \inst{2,3}
    \and
    A. de Koter
    \inst{2,3}
    \and
    H. Sana
    \inst{3}
    \and
    J. Poorta
    \inst{2}
    \and
    L. Kaper
    \inst{2}
    \and
    M. Stoop
    \inst{2}
          }

   \institute{
        Max Planck Institute for Astronomy, Königstuhl 17, 
            D-69117 Heidelberg, Germany\\
     \email{ramirez@mpia.de}
     \and
     Anton Pannekoek Institute for Astronomy, University of Amsterdam, Science Park 904, 1098 XH Amsterdam, The Netherlands
    \and
    Instituut voor Sterrenkunde, KU Leuven, Celestijnenlaan 200D bus 2401,
   	3001 Leuven, Belgium
             }

   \date{Received September 15, 1996; accepted March 16, 1997}

% \abstract{}{}{}{}{} 
% 5 {} token are mandatory
 
  \abstract
   {Significant progress has been made toward understanding the formation of 
   massive ($M>8$\,\msun) binaries in close orbits (with periods of less than a month). Some of the observational studies leading to this progress are the detection of a very low velocity dispersion among the massive stars in the young region M17 and the measurement of a positive trend of velocity dispersion with age in Galactic clusters. The velocity dispersion observed in M17 could be explained either by the lack of binaries among the stars in this region, which implies the highly unlikely scenario of a different formation mechanism for M17 than for other Galactic regions, or by larger binary separations than typically observed, but with a binary fraction similar to other young Galactic clusters.
   The latter implies that over time, the binary components migrate toward each other. This is in agreement with the finding that the radial velocity dispersion of young Galactic clusters correlates positively with their age.}
   {We aim to determine the origin of the strikingly low velocity dispersion by determining the observed and intrinsic binary fraction of massive stars in M17 through multi-epoch spectroscopy.}
   {We performed a multi-epoch spectroscopic survey consisting of three epochs separated by days and months, respectively. We complement this survey with existing data covering timescales of years. We determine the radial velocity of each star at each epoch by fitting the stellar absorption profiles. The velocity shifts between epochs are used to determine whether a close companion is present. }
   {We determine an observed binary fraction of 27\% and an intrinsic binary fraction of \intrFbin\%, consistent with that of other Galactic clusters. We conclude that the low velocity dispersion is due to a large separation among the young massive binaries in M17. Our result is in agreement with a migration scenario in which massive stars are born in binaries or higher order systems at large separation and harden within the first million years of evolution. Such an inward migration may either be driven by interaction with a remnant accretion disk, with other young stellar objects present in the system or by dynamical interactions within the cluster.
   Our results imply that possibly both dynamical interactions and binary evolution are key processes in the formation of gravitational wave sources.}
   {}


   \maketitle
%
%-------------------------------------------------------------------

\section{Introduction}

Multiplicity is a fundamental aspect of the process of massive star formation and evolution up until the formation of gravitational wave progenitors \citep[e.g.,][]{1991ASIC..342..125B, 2007ARA&A..45..481Z, 2012Sci...337..444S,  2012ARA&A..50..107L, 2013ApJ...764..166D, 2014prpl.conf..149T, 2023ASPC..534..275O}. The majority of massive stars (M\,>\,8\,M$_{\odot}$) are part of binary or higher-order multiple systems \citep{2014ApJS..215...15S, 2022A&A...663A..26B}, and about 40-50\% of these systems have an orbital period of less than a month. This is true in the Milky Way \citep[e.g.][]{2012Sci...337..444S, 2013A&A...550A.107S, 2015A&A...580A..93D, 2015ApJ...810...61M, 2017IAUS..329...89B, 2017ApJS..230...15M, 2022A&A...658A..69B} as well as in the lower metallicity environment of the LMC \citep[][]{2017A&A...598A..84A, 2021MNRAS.507.5348V}. Surveys of $\sim2-5$\,Myr old massive clusters reveal intrinsic spectroscopic OB binary fractions $f_{\rm bin}> 0.5$ and show multiple companions to be very common 
\citep[e.g.][]{1993PASP..105..588P,2006ApJ...639.1069H,2008MNRAS.386..447S, 2009A&A...507.1585R, 2009AJ....137.3437B, 2009AJ....137.3358M, 2009MNRAS.400.1479S, 2011MNRAS.416..817S,    2014ApJS..213...34K}. 
The orbital periods of these spectroscopically selected samples vary from about 1~day up to $\sim10$~years, with a preference towards close binaries with periods of less than one month (less than $\sim$1~au in separation). These are efficiently detected by spectroscopy through periodic Doppler shifts of the photospheric lines in their spectra \citep[e.g.,][]{2013A&A...550A.107S, 2015A&A...580A..93D, 2014ApJS..213...34K}.

The current understanding is that about 70\% of all O-type stars are so close, that they are expected to interact with their companion, about half of them before leaving the main sequence \citep[][]{2012Sci...337..444S}. About 25\% may even merge with a companion, offering a route towards the formation of the most massive stars \citep[e.g.][]{2005MNRAS.362..915B} and to formation of magnetic fields \citep[][]{2019Natur.574..211S, 2024Sci...384..214F}. Furthermore, massive binaries produce a variety of exotic products later in their evolution such as X-ray binaries, rare types of supernovae (Ibc, IIn), gamma-ray bursts, and, eventually, gravitational wave sources \citep{2011MNRAS.412.1522S, 1997ASPC..130..153P, 2016MNRAS.460.3545D, 2017PhRvL.119p1101A}.

\citet{2017A&A...604A..78R} and \cite{2017A&A...599L...9S} measured a distinctively low radial-velocity dispersion of 5.6$\pm$0.2\,\kms\ in a sample of 11 pre- and near-main sequence stars in the young (0.65$\pm$0.25\,Myr)
and nearest ($1675^{+19}_{-18}$\,pc)
giant H\,{\sc ii} region M17 \citep[][]{2024A&A...681A..21S}. The observed very low velocity dispersion could be explained by two scenarios: 1) The lack of binaries among the stars in M17, or 2) a binary fraction similar to those observed in other, somewhat older Galactic clusters, but larger binary separations. In order to determine which scenario is causing the low velocity dispersion in M17, we carried out a multi-epoch spectroscopic campaign of a larger number of stars to determine the binary fraction of the massive stars in this region. 

The targets in the sample used by \cite{2017A&A...599L...9S} (hereafter the 2017 sample) were selected based on the brightness of the K-band, and on the presence of emission lines in their K-band spectra presented by \cite{1997ApJ...489..698H}. The 2017 sample contains seven stars with infrared (IR) excess longwards of 3\,\micron, five of which also have a hot gaseous inner disk identified by double peaked emission lines and CO bandhead emission \citep{2017A&A...604A..78R}. 
The 2017 sample selection favors objects with a disk, which might potentially be a selection bias against binaries, since disk removal might be sped up under the influence of a companion star \citep[][]{1996ApJ...458..312J, 2012ApJ...745...19K, 2015ApJ...813...83C}.
Therefore, in this study, we changed the strategy of source selection to contain the most massive stars in M17 from the study by \citet{2013ApJS..209...32B}. Additionally, we included the stars from the previous study as to re-investigate the strikingly low velocity dispersion and assess possible biases.

In this paper, we first discuss the observed sample and radial-velocity measurements (Section~\ref{P5:sec:obs and data red}). We continue with the radial-velocity dispersion from a single-epoch analysis, and the observed and intrinsic binary fractions which follow from a multi-epoch analysis (Section~\ref{P5:sec:multiplicty analysis}). The results are discussed in Section~\ref{P5:sec:discussion}. Finally, our conclusions are presented in Section~\ref{P5:conclusions}.

%--------------------------------------------------------------------
\section{Observations and data reduction}\label{P5:sec:obs and data red}

A sample of 27 targets was chosen such that it includes the most massive stars in M17 \citep[from][]{2013ApJS..209...32B} as well as several pre-main sequence stars \citep[see][]{1997ApJ...489..698H, 2017A&A...604A..78R}. They all reside in the central cluster of M17 (NGC\,6618). 
The brightness of the objects ranges from $\sim$13 to $\sim$18~mag in the $V$ band and from $\sim$7 to $\sim$13~mag in the $K$ band; this large range is mainly due to the high and variable extinction to each of the sources in M17.  

In order to optimize the binary detection we designed the observing campaign such that each star is targeted three times: the first two epochs separated by at least 48 hours and the third one four weeks after. Adopting a flat mass-ratio distribution \citep{2012Sci...337..444S}, this approach provides binary detection rates of 70\% for periods up to 6 months, and better than 90\% for periods up to 10 days \citep[][]{2013A&A...550A.107S} and it allows us to obtain statistical constraints on their period distribution through the radial-velocity amplitude. However, due to bad weather conditions the proposed observing program could not be completed in full, therefore the number of observations varies from one to five epochs per star including the 2017 \xsh\ sample from \citet{2017A&A...604A..78R} (see Table~\ref{P5:tab:sample-overview}). 

\subsection{Data Reduction}

The 2019 VLT/\xsh\ \citep[][]{2011A26A...536A.105V} observations were taken between May and September under good weather conditions with seeing ranging between 0.5\arcsec\ and 1.2\arcsec\ and clear sky. The slit widths used were 1\arcsec, 0.9\arcsec, and 0.6\arcsec\ corresponding to a spectral resolution of 5100, 8800, and 8100 for the UVB, VIS and NIR arms, respectively. All spectra were taken in nodding mode. For some stars additional epochs from archival observations are added to the analysis. The log of observations is shown in Appendix~\ref{P5:app:obslog}. 

The spectra were reduced using the ESO-pipeline \texttt{esorex~3.3.5} \citep{2010SPIE.7737E..28M} with the exception of the 2012 observation of B331, for which we used the reduced spectra from \citet{2017A&A...604A..78R}. The flux calibration is done with spectrophotometric standards from the ESO database. We used \texttt{Molecfit~1.5.9} to correct for the telluric lines \citep{2015A&A...576A..77S, 2015A&A...576A..78K}. For B243, B268, B275, and B337, where the nebular contamination is significant, the nebular subtraction was performed by reducing each nodding position in staring mode and fitting the non-sky subtracted nebular lines. The fitted nebular spectrum was then removed from the science spectrum as described in \citet{2020A&A...636A..54V, 2024A&A...681A.112D}. 

\subsection{Stellar classification}
\label{P5:sec:stellar_classification}
For most of the targets, we use the spectral classification of \citet{2017A&A...604A..78R} and \citet{2024A&A...681A.112D}. For the remaining stars the spectral type has been determined in this paper by comparing the stacked spectra with the LAMOST survey \citep{Liu2019}, Galactic O-Star Catalog \citep[GOSC, ][]{2011ApJS..193...24S} and Gray's Atlas \citep{gray2000digital}. All spectral types are reported in Table~\ref{P5:tab:sample-overview}. Targets 326 and B86 are identified as double-lined spectroscopic binaries (SB2; see Section~\ref{P5:sec:multiplicty analysis}). More details on the spectral classification can be found in Appendix~\ref{P5:app:spectral classification}.

\begin{table}[]
\setlength{\tabcolsep}{3pt}
\caption{Sample overview. The first column shows the targets included in the \citet{2017A&A...599L...9S} sample and the new targets from the last observing campaign separated in two groups: the O and B stars included in the analysis of this paper and the targets identified as low-mass stars. The second and third columns list the name and spectral type, column 4 shows the number of observations for each target and the last column indicates whether or not the stars have IR-excess. Targets in bold face classify as binaries following the analysis in Section~\ref{P5:sec:observed_fraction}. Targets 326 and B86 are double-lined spectroscopic binaries (SB2).}
\begin{tabular}{cllcc}
\toprule
\multicolumn{1}{l}{}              & Star  & Sp. T.             & N epochs & IR excess \\ \midrule \midrule
\multirow{11}{*}{RT17, Sana17}    & B111  & O4.5\,V         & 4          &           \\
                                  & B164  & O6\,Vz          & 4      &           \\
                                  & B215  & B1\,V           & 3      & yes       \\
                                  & B243  & B9\,III         & 5       & yes       \\
                                  & \textbf{B253}  & B3-B5\,III      & 3      &           \\
                                  & B268  & B9\,III         & 5      & yes       \\
                                  & B275  & B7\,III         & 3      & yes       \\
                                  & B289  & O9.7\,V         & 4       & yes       \\
                                  & B311  & O8.5\,Vz        & 4        &           \\
                                  & B331  & late-B          & 1     & yes       \\
                                  & B337  & late-B          & 3        & yes       \\ \midrule
\multirow{10}{*}{New OB}          & \textbf{326}   & O6\,III + B1\,V & 3        &           \\
                                  & \textbf{B86}   & B9\,III + ?        & 3      &           \\
                                  & B93   & B2\,V           & 2      &           \\
                                  & \textbf{B150}  & B5\,V           & 3        & yes       \\
                                  & B181  & O9.7\,III       & 1          &           \\
                                  & B205  & B2\,V           & 3        &           \\
                                  & B213  & B3\,III-V       & 1        &           \\
                                  & B234  & B3\,III-V       & 1         &           \\
                                  & B272  & B7\,III         & 1          &           \\
                                  & CEN55 & B2\,III         & 1        &           \\ \midrule
\multirow{6}{*}{Low mass}       & B228  & G-type star     & 3         &           \\
                                  & B230  & F-type star     & 3         &           \\
                                  & B269  & F-type star     & 2       &           \\
                                  & B290  & A-type star     & 3       &           \\
                                  & B293  & F-type star     & 2      &           \\
                                  & B336  & G-type star     & 3      &           \\ \bottomrule
\end{tabular}
\label{P5:tab:sample-overview}
\end{table}

\subsection{Sample}
As we are interested in the binary fraction of the massive stars in M17, we remove from the sample B228, B230, B269, B290, B293 and B336 which are cool stars (A-, F- or G-type) and probably in the foreground of M17 based on their Gaia distance measurements. This leaves us with a sample of 21 OB stars with one to five epochs of observation. From these 21 stars, eight stars show an IR excess in their spectral energy distribution \citep[SED;][\textit{in press}]{Backs24}. 
Seven of these stars with disks are part of the 2017 sample. 
The masses of our targets range from 3 to 30~\msun\ \citep[][\textit{in press}]{Backs24}. 
Six stars (B181, B213, B234, B272, B331 and CEN55) only have one observation preventing us from calculating radial-velocity variations between spectra. This leaves us with a multi-epoch sample size of 15 stars. The single-epoch spectra can still be used to determine the radial-velocity dispersion of the sample (see Sec~\ref{P5:sec:RV_dispersion}).

\subsection{Radial-velocity measurements}
\label{P5:sec:RV_measurements}

We measured the radial velocity (RV) of the stars in each epoch (including the five stars with only one epoch) following the procedure described by \citet{2013A&A...550A.107S}. We developed a python code ({\sc RVFitter}\footnote{\url{https://github.com/WolfXeHD/RVFitter} \\ \url{https://github.com/MaclaRamirez/RVFitter_scripts}}) that allows to fit a given set of spectral lines using Gaussian, Lorenzian or Voigt profiles. 
The outcome of {\sc RVFitter} are two velocity measurements: i) the RV resulting from fitting all lines individually, allowing each line per epoch and star to have a different line depth and width (\textit{\_\_without\_constraints\_\_}), where the RV is given by the median of the measurements for all lines in a given epoch and the error corresponds to the standard deviation. The second output (\textit{\_\_with\_constraints\_\_}) is the RV assuming that all lines at a given epoch share the same radial velocity and that a given line fit has a constant amplitude and standard deviation as assumed in \citet{2013A&A...550A.107S}. 
In the latter case, the error is estimated from the covariance matrix. 
For this paper we adopt the velocities given by the method \textit{\_\_with\_constraints\_\_} and used Gaussian profiles for the fit with the exception of B181, B243, B268, B337 and B293, where we only fit hydrogen lines to determine RV and therefore use Lorenzian\footnote{Using a Voigt profile yields the same result as using a Lorenzian profile.} profile
\textit{\_\_with\_constraints\_\_}. 
The results of adopting Gaussian and Lorenzian profiles agree within the 1$\sigma$ errors.  
The RV values of each star per epoch can be found in Table~\ref{P5:tab:RV_measurements} and the lines used per star are given in Appendix~\ref{P5:app:lines_used}. We also measured the RVs of lines originating from the interstellar medium (ISM) to check and correct the RVs of the stars to account for differences in wavelength calibration between the epochs. 
The typical error on the RV measurements is about 0.2\,\kms\ and the differences between the RVs of the ISM lines are of the order of $1-3$\,\kms, except for B268 and B337 which have differences of $\sim$8\,\kms. 
Source 326 is a similar mass SB2 with overlapping absorption features. Therefore, the RVs could not be measured for this star. B86 is also an SB2, but the mass and luminosity difference between both stars is large enough that a few single absorption lines of one of the stars could be used to obtain RV measurements.

\section{Multiplicity analysis}\label{P5:sec:multiplicty analysis}

In this section we discuss the analysis of the multiplicity of the stars in our sample. In Section~\ref{P5:sec:RV_dispersion} we present the RV dispersion (\srv) of the newly obtained sample as if we had single-epoch observations in order to compare our results to the work presented in \citet{2017A&A...599L...9S} and \citet{2021A&A...645L..10R}. In Section~\ref{P5:sec:observed_fraction} we establish the spectroscopic binary fraction based on the radial-velocity variations (SB1) and/or the SB2 nature of the spectra and Section~\ref{P5:sec:intrinsic_fbin} establishes the intrinsic binary fraction in M17. 

\subsection{Radial-velocity dispersion}\label{P5:sec:RV_dispersion}

\citet{2017A&A...599L...9S} and \citet{2021A&A...645L..10R} performed Monte Carlo simulations in order to find the minimum orbital period (\pcutoff) and binary fraction (\fbin) that best reproduced the observed \srv\ measured in clusters with only single-epoch observations. 
In order to compare our newly obtained multi-epoch data to the single-epoch data represented in \citet{2021A&A...645L..10R} we followed the procedure described by the authors. In short, we computed the most probable \srv\ for each cluster by randomly drawing
one epoch per star and computing the RV dispersion \srv\ of the sample cluster as:

\begin{equation}
\sigma_{\rm 1D} = \sqrt{\dfrac{\sum_{i=1}^{n} w_i \cdot ({\rm RV}_i - \langle v_{\rm RV}\rangle)^2}{\sum_{i=1}^{n} w_i}},~\mathrm{where} \\
\langle v_{\rm RV}\rangle= \frac{\sum_{i=1}^{n} w_i {\rm RV}_i}{\sum_{i=1}^{n} w_i}  
\label{P5:eq:s1d}
\end{equation}

\noindent and ${\rm RV}_i$ is the radial-velocity of a star at a certain epoch given by ${{\rm RV_{star, \textit{i}} - RV_{ISM, \textit{i}}}}$ (${\rm RV_{star}}$ and $\rm {RV_{ISM}}$ are given in Table~\ref{P5:tab:RV_measurements}), $\langle v_{\rm RV}\rangle$ is the weighted mean radial velocity (given in Table~\ref{P5:tab:Cluster_properties}), and $w_i=1/\sigma_{\rm RV, \textit{i}}^{2}$ the weight for each measurement.
We repeated this procedure 10,000 times and then calculated the most probable \srv\ as the 50$^{\rm{th}}$ percentile. The corresponding uncertainty is taken as half of the difference between the 16$^{\rm{th}}$ and 84$^{\rm{th}}$ percentiles of the resulting \srv\ distributions. 
The error on the distribution is slightly higher than those presented in \citet{2021A&A...645L..10R}. This difference is due to the fact that in this paper we calculate the error in \srv\ based on the percentiles and not as the standard deviation of a Gaussian fit to the \srv\ distribution. The latter is to avoid making the assumption that the \srv\ distributions are Gaussian. 
Using the spectra of \citet{2017A&A...599L...9S}, but with the analysis presented here, we find a \srv\ of \srvMorig\,\kms. This is slightly higher than reported in 2017 ($5.6\pm0.2$\,\kms), but still very low in comparison to slightly older Galactic clusters. The difference is due to the new reduction, normalization and clipping of the data. 

With the extended sample presented in this paper we obtain a \srv\ for M17 of \srvM\,\kms, which is higher than the $5.6\pm0.2$\,\kms\ reported for the 2017 sample. The difference is due to the inclusion of three stars with significant RV variations (see Section~\ref{P5:sec:discussion}). However, these values remain much lower than the than the observed \srv\ of $\sim30$\,\kms\ in other young ($\sim2-5$\,Myr) clusters with significant short period binary populations.

\begin{figure}
\includegraphics[width=0.93\hsize]{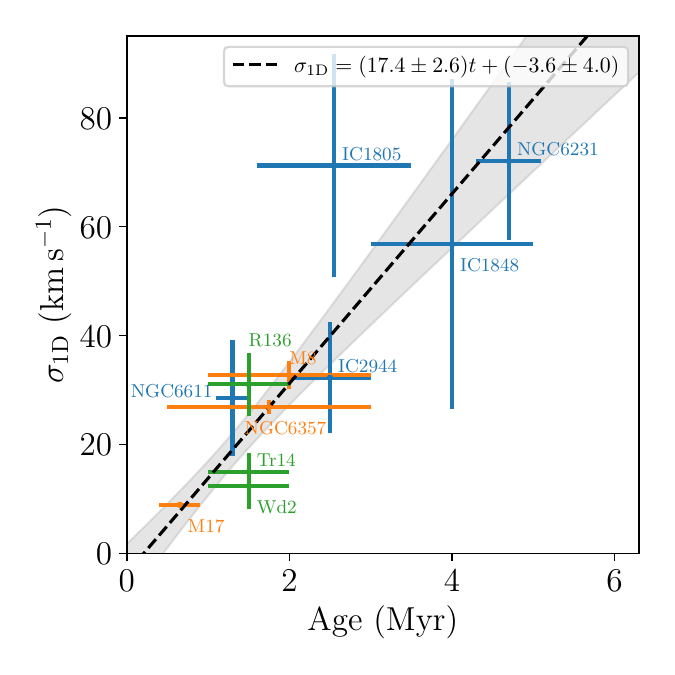}
  \caption{Observed radial-velocity dispersion \srv\ as a function of age of a number of Galactic clusters and R136 in the LMC. The dashed black line shows the positive correlation of \srv\ with age \citep[see also][]{2021A&A...645L..10R}. The gray shaded region indicates the 1$\sigma$ uncertainty of the fit. Blue points indicate the clusters included in \citet{2012Sci...337..444S}, green points show Wd2, Trumpler~14 and R136 and orange points M17 and the young clusters in \citet{2020A&A...633A.155R}.}
     \label{P5:fig:RVdisp_vs_age}
\end{figure}

With this newly calculated \srv\ and updated age estimates for NGC\,6611 from \citet{2023A&A...670A.108S}, NGC\,6231 from \citet{2021A&A...655A..31V}, and M17 from \citet{2024A&A...681A..21S}, we can update the \srv\ vs age relation first introduced in \citet{2021A&A...645L..10R}. 
To calculate \srv\ for the clusters with more than one epoch observations we followed the procedure described in \citet{2021A&A...645L..10R}. We drew RV values for each star at a random epoch from a Gaussian centered at the reported RV and with a width equal to the measurement uncertainty. We repeated this procedure $10^5$ times and then calculated the most probable \srv\ and its standard deviation.
The updated relation is shown in Figure~\ref{P5:fig:RVdisp_vs_age} and the data is presented in Table~\ref{P5:tab:Cluster_properties}: we include the clusters studied by \citet{2012Sci...337..444S}, those by \citet{2020A&A...633A.155R}, R136 in the Large Magellanic Cloud (LMC) from \citet{2012AA...546A..73H}, Westerlund~2 (Wd2) from \citet{2018AJ....156..211Z} and Trumpler~14 from \citet{2018MNRAS.477.2068K}. 
With the updated data, after performing an orthogonal distance regression (ODR), we find a positive correlation between \srv\ and the age of clusters. The Pearson coefficient is 0.8, which indicates a strong correlation, and is higher than that found by \citet{2021A&A...645L..10R}.
We note that with the exception of NGC~6611, the stars in 0 to 2~Myr old clusters span a lower mass range than the more massive clusters (see Table~\ref{P5:tab:Cluster_properties}). This could partially explain the low \srv\ as the binary properties vary with primary mass \citep[e.g.][]{2017ApJS..230...15M}. Nevertheless, restricting the samples of M17 and Wd2 to stars with $M>15$~\msun\ does not change our conclusions.

\subsection{Observed spectroscopic binary fraction}
\label{P5:sec:observed_fraction}

In order to determine the spectroscopic binary fraction we perform a statistical test as described in \citet{2013A&A...550A.107S}. 
A star is detected as a binary if the following two criteria are fulfilled simultaneously: a) two measurements deviate significantly from each other, and b) those same measurements differ by a minimum threshold \textit{C}. The latter condition is to avoid RV variations caused by pulsations or photospheric variability. To make this concrete, a star is a binary if it meets the following criteria:

\begin{align}
\centering
\frac{|\mathrm{RV}_i - \mathrm{RV}_j|}{\sqrt{\sigma_{\mathrm{RV}, i}^2 + \sigma_{\mathrm{RV},j}^2}} > 4.0 && \mathrm{and} &&  |\mathrm{RV}_i - \mathrm{RV}_j| > 20~{\rm km\,s}^{-1},
\end{align}

\noindent where RV and $\sigma_{\mathrm{RV}}$ correspond to the RV measured and its corresponding one sigma error at a given epoch. As stated in \citet{2013A&A...550A.107S}, the confidence threshold of 4.0 gives a false positive rate of 1/1000 given the sampling of our measurements. The value of $C=20$~\kms, correspond to the maximum RV variation that can be caused by stellar variability \citep[e.g.][]{2009A&A...507.1585R}. 

Following this method, we detect three sources (B86, B150 and B253) with significant RV variations in our sample with multi-epoch observations. Including source 326, which is an SB2, we find a total of four binaries in our sample of 15 stars with multi-epoch spectroscopy. 
The latter results in an observed spectroscopic binary fraction \fbin\,$=$\,\obsFbin, where the $1\sigma$ error comes from the size of the sample. 
Adopting $C$ values of 15 and 10\,\kms\ would result in an observed spectroscopic binary fraction of our sample of \fbin\,$=0.33\pm0.12$ and \fbin\,$=0.40\pm0.13$, respectively, the difference being that lower $C$ values are more favorable to detect longer period binaries.

Including the two SB2s in the center of M17 \citep[CEN1a and CEN1b;][see Figure~\ref{P5:fig:spat_corr}]{2022A&A...663A..26B}, we obtain an observed binary fraction \fbin\,$=0.35\pm0.08$. A recent study from \citet{2024A&A...681A..21S} shows that about 30\% of the O stars of the original population in the center of M17 have been dynamically ejected. In total they find 13 OB runaway stars. From single epoch spectroscopy, two of those runaways are identified as SB2s. Assuming that the rest of their sample consists of single stars, including the SB2 runaway stars results in an observed binary fraction \fbin\,$=0.21\pm0.08$ (8 binaries out of 30 stars). We note that this value corresponds to a lower limit of the observed \fbin, as it is possible that other runaway stars are also binaries. Multi-epoch spectroscopy is needed in order to better constrain the observed \fbin\ including the runaways. 

\subsection{Intrinsic binary fraction}
\label{P5:sec:intrinsic_fbin}

The binary fraction reported in Section~\ref{P5:sec:observed_fraction} corresponds to a minimum estimate of the real binary fraction of the OB stars in M17. The intrinsic fraction depends on the sensitivity of our observations in terms of time sampling and RV measurement accuracy as well as on the orbital properties and orbital inclinations of the stars in M17. In order to provide constraints on the intrinsic binary fraction of our sample we take a Monte-Carlo approach. The methodology is based on that described in \citet{2013A&A...550A.107S}.

In this approach we simulate multi-epoch RV measurements of stars in clusters and determine the largest change in RV ($\Delta$RV) for the individual stars. The distribution of $\Delta$RV values depends on the time sampling, which is chosen identical to the observations, and the intrinsic binarity of the population and orbital properties of the binaries. We determine the agreement between the simulated and the observed $\Delta$RV distributions using a Kolmogorov-Smirnov (KS) test. We vary \fbin\ and \pcutoff\ considered in the simulations to find a best fitting simulated distribution. Below we list the distributions considered for the other orbital parameters. 

For the primary masses, we adopt the masses derived in Section~\ref{P5:sec:stellar_classification}, and create random systems drawing from distributions based on the binary properties derived by \citet{2012Sci...337..444S} for Galactic young clusters. 
The mass-ratio distribution is uniform with $0.1 < q < 1$. The probability density function of the period is described as $\text{pdf}(P) \propto (\log P)^{-0.5}$, with the period in days and $\log P_{\rm cutoff} < \log P < 3.5$. The eccentricity distribution depends on the period of the binary system: for $P<4$ days, we assume circular orbits; for periods between 4 and 6 days, the eccentricities are sampled from $\text{pdf}(e) \propto e^{-0.5}$, with $0 \leq e < 0.5$; for periods longer than 6 days the same distribution is used, but with $0 \leq e < 0.9$. We vary the binary fraction \fbin\ from 0.1 to 1 in steps of 0.01 and the \pcutoff\ from 1.4 to 3500~days in steps of 1~day. 

\begin{figure}[ht!]
\includegraphics[width=\hsize]{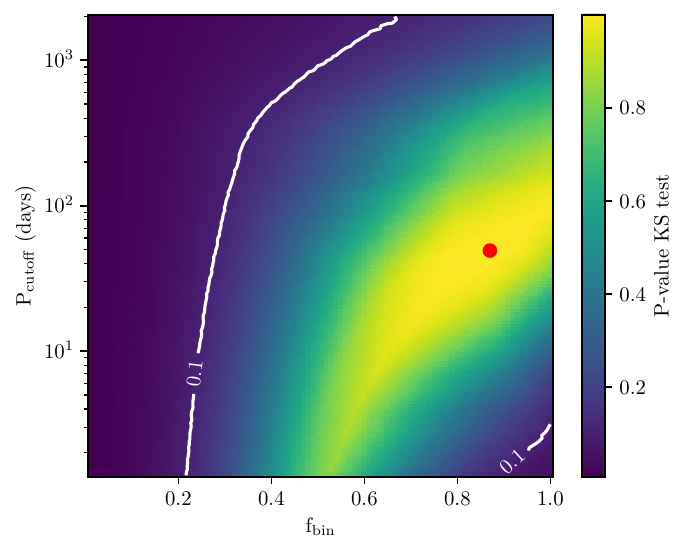}
  \caption{$p$-values (color bar) for the KS-test obtained from comparing the $\Delta$RV from our observations to that of simulated populations with different combinations of \fbin\ and \pcutoff. The white line indicates the 0.1 contour outside which we can reject the null hypothesis that the simulated populations are drawn from the same parent population as our observations. The red dot at \fbin\,$=$\,\intrFbin\% and \pcutoff\,$=$\,\intrPcutoff\,days shows the population that best represents our observations.}
     \label{P5:fig:kstest}
\end{figure}

Figure\,\ref{P5:fig:kstest} shows the resulting $p$-values for the KS test in the above described procedure for different combinations of \fbin\ and \pcutoff. The figure shows the general trend that our observations can be drawn from parent populations containing wider orbits as the binary fraction increases.   
We find that our observed sample has the highest probability to have been drawn from a parent population with a binary fraction of \intrFbin\% and a minimum period of \intrPcutoff~days. 
The size of the errors on the RV measurements has a significant influence on the results. 
Increasing the RV errors by a few \kms\ has the consequence that populations with lower intrinsic binary fractions have a higher probability of being drawn from the same parent distribution as our observations. 
The contour in Figure~\ref{P5:fig:kstest} shows $p$-values from the KS-test of 0.1. Only populations outside that contour can be rejected as being drawn from the same parent population as our sample.

\section{Discussion}\label{P5:sec:discussion}

\subsection{Radial-velocity dispersion of M17}

\begin{figure}
\centering
\includegraphics[width=\hsize]{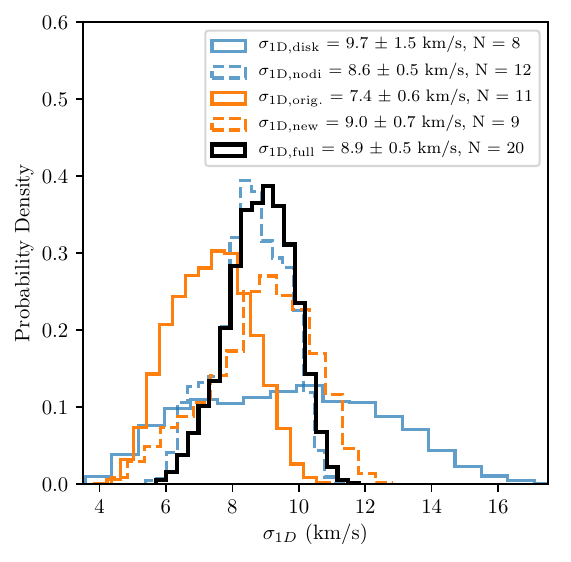}
  \caption{\srv\ for our full sample and for sub-samples thereof. The full sample is shown with the solid black line. The orange lines show the 11 stars from \citet{2017A&A...599L...9S} (dashed line) and the nine new stars added to the sample in this paper (solid line). The blue lines show the sample divided in stars with (solid line) and without disks (dashed line).}
     \label{P5:fig:RVdisp_M17}
\end{figure}

In order to explore possible biases in the original target selection, in this section we explore the effect of measuring \srv\ on different sub-samples of our M17 dataset. Figure~\ref{P5:fig:RVdisp_M17} shows the \srv\ distribution calculated in Section~\ref{P5:sec:RV_dispersion} for our sample of young stars in M17 with the exception of source 326 (SB2), for which we could not measure RV ($N=20$), and for two sets of sub-samples; the stars added in this work (new stars; $N=9$) vs the original sample \citep[$N=11$;][]{2017A&A...599L...9S, 2021A&A...645L..10R} and the stars with IR excess ($N=8$) vs those with no IR-excess \citep[$N=12$;][\textit{in press}]{Backs24}. 
On the one hand, the \srv\ obtained for the stars with (solid blue line) and without (dashed blue line) disks agree within the errors. 
Nevertheless, the shapes of the distributions differ significantly. The latter is because of the presence of the binary B150 in the disk sample; this star shows significant radial velocity variations that cause the distribution to broaden. Another star with significant $\Delta$RVs is present in the sample without disks, but in this case the sample size is larger, so the effect of including it is not as significant as in the sample with disks. 
That the \srv\ in both sub-samples is similar suggests that in M17, which does not host many close binaries, the lifetime of disks is not affected due to binary interactions. 
On the other hand, the distributions showing the original 2017 sample (dashed orange line) vs the newly added stars (solid orange line), show a significant difference in \srv, where the original sample has a lower \srv. 
Nevertheless, these two sub-samples are consistent with each other considering their 1.5 sigma uncertainty. 
If the difference was real, this would imply that the original sample selected by \citet{2017A&A...604A..78R} lacks close binaries with respect to the sample in this paper (including the binaries B86, B150 and B253). 
However, given the newly included binaries in both the distributions with and without disks, we do not think that the reason for the difference in the original vs new sample is due to a bias in the 2017 sample selection, which was tailored to target young objects with circumstellar disks. 

\subsection{Spatial correlation}

\begin{figure}
\includegraphics[width=\hsize]{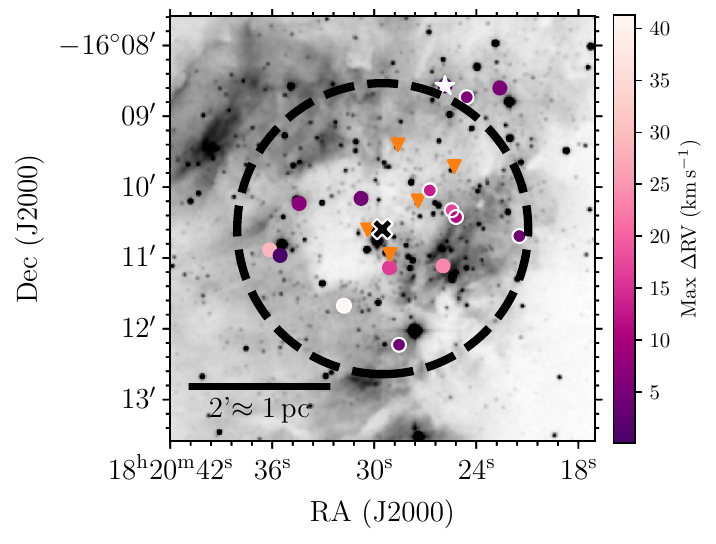}
  \caption{Location of our targets with respect to the center of M17. The dashed circle shows the size of the cluster \citep[$\sim2$\arcmin;][]{2024A&A...681A..21S}, centered on the position of CEN\,1ab  (black cross). The color bar represents the maximum radial velocity difference between any two epochs for the sources with multi-epoch spectra. The sources with disks are marked with white edges and the SB2 326 is shown with a white star. The position of the sources with single epoch observations is shown with an orange triangle. The figure shows there are no indications that the location of the shortest-period binaries is dependent on distance to the cluster center.}
     \label{P5:fig:spat_corr}
\end{figure}

Figure~\ref{P5:fig:spat_corr} shows the position of our O and B stars with respect to the center of NGC\,6618 \citep[18:20:29.53, -16:10:35.56;][]{2024A&A...681A..21S}. The colors of the markers indicate the maximum radial-velocity difference $\Delta$RV between any two epochs, with the larger $\Delta$RVs corresponding to the close binaries. 
The location of the very high mass multiple system CEN\,1ab coincides with the location of the center of the cluster. 
There is no evident correlation with location of the stars displaying the largest $\Delta$RV. This suggests that close binarity in M17 is not dependent on the distance to the cluster center, and indicates that our result about the lack of close binaries in M17 is independent of the position of the selected targets.

\subsection{Transversal velocity dispersion}

The radial velocities measured in this paper represent the motion of individual stars along the line of sight and are usually dominated by binary motions (or intrinsic variability such as pulsations). 
In contrast, the proper motions reflect the velocity of the systems inside the cluster and are a measurement of their movement due to the cluster dynamics. 
With the aim of comparing these two quantities, in this section we derive the 1D velocity dispersion of the clusters in Figure~\ref{P5:fig:RVdisp_vs_age} based on the proper motion of the cluster members ($\sigma_{\rm 1D, RA}$, $\sigma_{\rm 1D, DEC}$). 
To find the cluster members, we used the cluster catalog with Gaia DR3 by \citet{2023A&A...673A.114H}. We exclude R136, because being in the LMC it is not included in their catalog. 
To calculate $\sigma_{\rm 1D, RA}$, $\sigma_{\rm 1D, DEC}$ we used Equation~\ref{P5:eq:s1d} and the distance listed in \citet{2023A&A...673A.114H} for the clusters with more than 100 listed members. 
For the three clusters in which the catalog did not have enough stars, M17, IC\,2944 and Wd2, we adopted distances from \citet{2024A&A...681A..21S}, \citet{2023ApJS..265...59Z}, \citet{2018A&A...618A..93C}, respectively. 
Table~\ref{P5:tab:Cluster_properties} lists $\sigma_{\rm 1D, RA}$, $\sigma_{\rm 1D, DEC}$ and the adopted distance for each of the clusters. 
For all clusters these values range from 0.4 to 2.3\,\kms\ and do not show a relation with the age of the cluster.  
This is in agreement with the migration scenario as we expect the binaries to harden, and therefore \srv\ to increase with age, but the dispersion of the cluster due to cluster dynamics ($\sigma_{\rm 1D, RA}$ and $\sigma_{\rm 1D, DEC}$) to not vary significantly within the explored time frame (0 to $\sim5$~Myrs). Additionally, it shows that typical cluster motions alone cannot explain the measured \srv\ and implies that there must be a contribution from binary motions. 

The latter underlines the need to be careful when using \srv\ to derive virial masses of star clusters. For the young clusters studied here the line-of-sight velocity dispersion is dominated by binary motions rather than by random motions of the individual systems, therefore the cluster mass is much less than implied by the virial mass (proportional to \srv$^{2}$). Cluster masses of star clusters in virial equilibrium can be constrained reliably using the dispersion of the transverse velocity component.

\begin{table*}[h]
\caption{Properties of the sample of young clusters plotted in Figures~\ref{P5:fig:RVdisp_vs_age} and \ref{P5:fig:Pcutoff_vs_age}.}             % title of Table
\label{P5:tab:Cluster_properties}      % is used to refer this table in the text
\centering                          % used for centering table
\renewcommand{\arraystretch}{1.4}
\setlength{\tabcolsep}{3pt}
\begin{tabular}{lccccccccccc}
\hline
Cluster & Age & distance & $\sigma_{\rm 1D}$ & $\sigma_{\rm 1D, RA}$ & $\sigma_{\rm 1D, DEC}$ & $\langle v_{\rm RV}\rangle$ & N & Mass & \pcutoff & Age & Dist. \\
 & Myr & pc & \kms\ & \kms\ & \kms\ & \kms\ & stars & \msun & days &  ref. & ref.\\
\hline
IC1805 & 1.6 -- 3.5 & $2019^{+6}_{-6}$ & $71.2\pm20.6$ & $1.44\pm0.01$ & $2.25\pm0.01$ & $-53.6\pm24.0$ & 8 & 15 -- 60 & $1.4_{-0.0}^{+0.8}$ & 1 & 10 \\
IC1848 & 3.0 -- 5.0 & $2058^{+8}_{-8}$ & $56.8\pm30.3$ & $1.25\pm0.02$ & $1.01\pm0.02$ & $-42.2\pm28.1$ & 5 & 15 -- 60 & $1.4_{-0.0}^{+1.7}$ & 2 & 10 \\
IC2944 & 2.0 -- 3.0 & $2138_{-203}^{+210}$ & $32.3\pm10.2$ & $0.40\pm0.04$ & $0.52\pm0.04$ & $-0.4\pm8.6$ & 14 & 15 -- 60 & $6.3_{-4.5}^{+16.0}$ & 3 & 11 \\
NGC6231 & 4.3 -- 5.1 & $1553^{+2}_{-2}$ & $72.0\pm14.6$ & $1.07\pm0.01$ & $1.14\pm0.01$ & $-14.2\pm21.2$ & 13 & 15 -- 60 & $1.4_{-0.0}^{+0.4}$ & 4 & 10 \\
NGC6611 & 1.1 -- 1.5 & $1698^{+5}_{-5}$ & $28.5\pm10.6$ & $1.58\pm0.01$ & $1.93\pm0.01$ & $16.8\pm9.0$ & 9 & 15 -- 60 & $6.8_{-5.4}^{+29.0}$ & 5 & 10 \\
Wd2 & 1.0 -- 2.0 & $4208^{+74}_{-80}$ & $12.4\pm4.2$ & $2.48\pm0.09$ & $2.13\pm0.08$ & $8.4\pm2.2$ & 44 & 6 -- 60 & $203.4_{-124.6}^{+281.8}$ & 6 & 12 \\
M8 & 1.0 -- 3.0 & $1229^{+3}_{-3}$ & $32.7\pm2.6^{(*)}$ & $2.26\pm0.01$ & $1.46\pm0.01$ & $-2.5\pm2.5^{(*)}$ & 16 & 6 -- 20 & $2.4_{-1.0}^{+4.9}$ & 7 & 10 \\
NGC6357 & 0.5 -- 3.0 & $1672^{+7}_{-7}$ & $26.9\pm1.3^{(*)}$ & $1.07\pm0.01$ & $1.14\pm0.01$ & $16.6\pm1.3^{(*)}$ & 22 & 6 -- 30 & $5.4_{-3.3}^{+9.6}$ & 7 & 10 \\
M17 & 0.4 -- 0.9 & $1675^{+19}_{-18}$ & $8.9\pm0.5$ & $1.47\pm0.03$ & $2.20\pm0.02$ & $9.5\pm0.7$ & 20 & 3 -- 50 & $414.3_{-329.0}^{+1764.0}$ & 9  & 9\\
R136 & 1.0 -- 2.0 & $49.59\times10^3$ & $31.0\pm5.8$ & $...$ & $...$ & $270.6\pm3.5$ & 332 & 15 -- 60 & $11.8_{-3.2}^{+4.4}$ & 8 & 13 \\
Tr~14 & 1.0 -- 2.0 & $2362^{+54}_{-54}$ & $14.9\pm3.5$ & $1.01\pm0.02$ & $0.85\pm0.02$ & $4.0\pm1.0$ & 32 & 15 -- 50 & $188.0_{-125.8}^{+297.3}$ & 14,15 & 16 \\
\hline
\end{tabular}
\tablebib{
(*) \srv\ and $\langle v_{\rm RV}\rangle$ from~\citet{2020A&A...633A.155R}, (1)~\citet{2017ApJS..230....3S}; (2)~\citet{2014MNRAS.438.1451L}; (3)~\citet{2014MNRAS.443..411B}; (4)~\citet{2021A&A...655A..31V}; (5)~\citet{2023A&A...670A.108S}; (6)~\citet{2018AJ....156..211Z}; (7)~\citet{2020A&A...633A.155R}; (8)~\citet{2012AA...546A..73H}; (9)~\citet{2024A&A...681A..21S};
(10)~\citet{2023A&A...673A.114H}; (11)~\citet{2023ApJS..265...59Z};
(12)~\citet{2018A&A...618A..93C}; (13)~\citet{2019Natur.567..200P};
(14)~\citet{2010A&A...515A..26S}; (15)~\citet{2011MNRAS.418..949R};
(16)~\citet{2022A&A...660A..11G}
}
\end{table*}

\subsection{Eccentricity}

\cite{2015ApJ...810...61M} measured eccentricities of a sample of young B-type binaries and show that the youngest binaries (<\,1\,Myr) in the sample have eccentricities of $\sim0.6$. 
It could be argued that a sample with many binaries in eccentric orbits would result in a lower \srv, since the stars spend most of their time near apastron. 

In Appendix\,\ref{P5:app:eccentricity} we explore the effect of changing the eccentricity on the properties of the simulated populations. 
First we simulate single-epoch observations of populations with all properties as described in Section\,\ref{P5:sec:intrinsic_fbin}, and we vary the power-law index of the eccentricity sampling from -0.5 (corresponding to the \citet{2012Sci...337..444S} empirical value) to 3 (strong preference for eccentric orbits). 
Populations following the \citet{2012Sci...337..444S} empirical properties (for $\sim2-5$\,Myr old clusters) have \srv\ of around 40\,\kms. The effect of having preferentially eccentric orbits, even in the most extreme case where we allow all orbits to be eccentric, reduces the \srv\ of the simulated populations by at most 6\,\kms. 
This is not enough to explain the low \srv\ observed in M17. 
Second, we simulate a sample for which multi-epoch observations are available with the same properties as ours. Assuming \fbin\,$=0.7$ and \pcutoff\,$=1.4$\,days \citep[][]{2012Sci...337..444S} and having a strong preference for eccentric orbits has the effect of increasing the predicted $\Delta$RVs, which is the opposite of what we observe in M17, where the $\Delta$RVs are lower than those in $\sim2$ to 5\,Myr old clusters.
We conclude that more eccentric orbits cannot explain the low \srv\ nor the observed $\Delta$RVs of M17. In the best case scenario the effect of high eccentricities is minor compared to the effect of altering \fbin\ and \pcutoff\ of the sample.   

\subsection{Timescale for binary hardening}

Assuming a universal binary fraction of \fbin\,$=0.7$ \citep[][]{2012Sci...337..444S}, we can estimate the \pcutoff\ that best represents the measured \srv\ for each cluster following the procedure described in \citet{2021A&A...645L..10R}.
A more accurate synthetic radial velocity distribution could be obtained by including a variable binary fraction that depends on the mass of the primary star. However, due to the relatively small size of most of our samples, we refrain from including this and assume a single binary fraction for all stars.

The best fitting distribution to explain the \srv\,$=$\,\srvM\,\kms\ observed for M17 corresponds to \pcutoff\,$=$\,\pcutoffM\,days, and distributions with \pcutoff\ $<$\,\pcutoffMoSig\ and $<$\,\pcutoffMtSig\,days can be rejected with 68\% and 95\% confidence, respectively.
Adopting an intrinsic binary fraction for the young stars in M17 of \intrFbin\% as derived in  Section~\ref{P5:sec:intrinsic_fbin} leads to a best fit distribution with \pcutoff\,$=$\,\pcutoffMfeighty\,days, and distributions with \pcutoff\, $<$\,\pcutoffMfeightyOsig\ and $<$\,\pcutoffMfeightyTsig\,days can be rejected with 68\% and 95\% confidence, respectively.
The upper limit for \pcutoff\ derived for $\sim2$ to 5\,Myr old clusters is a few to a few tens of days, i.e., an order of magnitude lower than that derived for M17. 

These statistical constraints still allow for a few systems with shorter periods to exist. In fact, systems with periods $< 10$~days have been detected in young clusters, for example some B-type stars in the \citet{2015ApJ...801..113M} sample have companions with $P = 3.0-8.5$~days and extreme mass ratios. Additionally, \citet{2018MNRAS.477.2068K} find a binary with $P = 9.4$~days located in the outskirts of Tr~14 (HDE~303312).

\begin{figure}
\centering
\includegraphics[width=0.91\hsize]{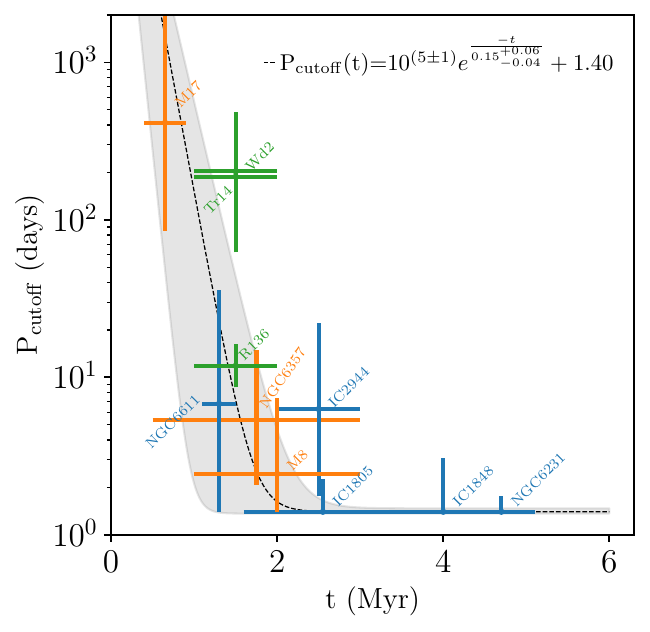}
  \caption{\pcutoff\ as a function of age of young clusters with \fbin$=0.70$. The uncertainties correspond to the distributions that represent the $16^{th}$ to $84^{th}$ percentile of the \srv\ distribution. The solid line and shaded region show the fit to the data and its 1$\sigma$ uncertainty.}
     \label{P5:fig:Pcutoff_vs_age}
\end{figure}

In order to get an idea of the timescale of binary hardening we fit the function $P(t) = P_{0}e^{-t/t_{0}} + c$ to the data in Figure~\ref{P5:fig:Pcutoff_vs_age}, where $P_0$ is the minimum period at the moment of binary formation and $t_0$ corresponds to the e-folding time. We assume a value for $P_0=10^5$\,days, which is the period corresponding to a pair of 10\,\msun\ stars with a separation of $\sim$100\,au \citep[][]{2021A&A...645L..10R} and $c=1.4$\,days, which is the cutoff period of $\sim2-5$\,Myr old Galactic clusters \citep{2012Sci...337..444S}.  
We obtain a typical e-folding time $t_0 =$\efoldtime\,Myr.
%$\frac{\Delta t}{t_0} = ln(P1) - ln(P2)
\citet{2012Sci...337..444S} finds the minimum period for $\sim2-5$\,Myr old clusters to be 1.4\,days. 
To harden an orbit from \pcutoffM\,days to 1.4\,days, ln(\pcutoffM/1.4)$\,\approx6$ e-folding times are needed. With $t_0=0.15$\,Myr the typical hardening timescale is $\sim$\,\hardtime\,Myr. 
For a system with two 10\,\msun\ stars, this would imply a hardening rate of \hardrate\,$\rm{au\,Myr^{-1}}$. 
Adopting a \pcutoff\ of \pcutoffMfeighty\,days would lead to a hardening timescale of \hardtimefeighty\,Myr, and a hardening rate of \hardratefeighty\,$\rm{au\,Myr^{-1}}$. 
Assuming \pcutoff\ of \pcutoffMoSig\ and \pcutoffMfeightyOsig\,days would lead to hardening rates of \hardrateOsig\,$\rm{au\,Myr^{-1}}$ and \hardratefeightyOsig\,$\rm{au\,Myr^{-1}}$, respectively.  
These rates are lower than the values of $4-7.4\,\rm{au\,Myr^{-1}}$ reported by \citet{2021A&A...645L..10R}; the reason for the difference is the significantly lower \srv\ measured in the 2017 sample, which leads to a much higher \pcutoff\ for the best fit distribution. Nevertheless, the hardening rates remain significant and require efficient dynamical processes.

\subsection{Migration scenario}

Both our single and multi-epoch analyses for a sample of 20 and 15 massive stars, respectively, in the young star-forming region M17 point to a lack of short period binaries relative to older clusters ($\sim2-5$\,Myr), where close binaries with periods of days are common. From the multi-epoch observations we derive a binary fraction that, within uncertainties, is consistent with the universal value of $\sim$70\% in massive clusters of ages up to $\sim10^{7}$\,yrs. These results support the migration scenario for close-binary formation as proposed by \citet{2017A&A...599L...9S} and \citet{2021A&A...645L..10R}.
We mention that if massive stars are preferentially in wider orbits during cluster formation this aids in the production of runaway stars, as the larger effective cross section of wide binary systems favors dynamical interactions \citep[e.g.,][]{2011Sci...334.1380F,2024A&A...681A..21S}.

Which physical process may be responsible for binary hardening? Mechanisms that have been proposed include interaction with a circum-binary disk,
the Eccentric Kozai-Lidov (EKL) effect in triple systems \citep{2017A&A...599L...9S}, dynamical interactions among stellar systems in the cluster environment and tidal circularization and orbital decay \citep[][]{2015ApJ...810...61M}. We discuss each of these possibilities below. Before doing so, we mention that dynamical disruption of co-planar (hierarchical) triples that initially fragmented within the disk in combination with energy dissipation occurring within the disk, may potentially harden the inner binary. However, this process is expected to be efficient in earlier pre-main sequence or proto-stellar phases \citep{2018ApJ...854...44M} and disk fragmentation is expected to produce predominantly low-mass companions \citep[][]{2020A&A...644A..41O} which are more difficult to detect spectroscopically.

\emph{Star-disk interaction:} For two reasons, we anticipate that interaction with a circum-binary disk is not an efficient mechanism for removal of orbital angular momentum, therefore hardening, of the binary system \citep[see also, e.g.,][]{2023arXiv231107278W}. First, 13 out of 21 sources in our sample do not show evidence for the presence of a disk, while the anticipated time period of binary hardening of $\sim$10$^{6}$\,yr is still ahead. 
This implies that in most cases disk removal happens prior to binary hardening, and may also suggest that the typical disk removal timescale is short compared to the binary hardening time.
%In other words, in many cases the characteristic disk removal time is short compared to the binary hardening time. 
Second, from ALMA mm-observations the total disk masses of B243, B268, B275, and B331 are estimated to be no higher than $9\times10^{-3}$\,\msun ~(Poorta et al., \emph{in prep.}). Radiation thermo-dynamical modeling of hydrogen emission lines of B243 and B331 yield disk masses of the order of $10^{-3}-10^{-4}$\,\msun\ inside 20\,au \citep{2023A&A...671A..13B}. It seems unlikely that these remnant disks are capable of storing the orbital angular momentum that is to be removed from the binary in the overall hardening process. 
The latter is in agreement with \citet{2020MNRAS.491.5158T}, who show that disk migration must occur during the embedded protostellar phase, which is much shorter than the age of our systems \citep[][]{2020MNRAS.491.5158T}.

\emph{Eccentric Kozai-Lidov mechanism:} From interferometric observations of a sample of six young O-type stars in M17 with VLT/GRAVITY, \citet{2022A&A...663A..26B} find a companion fraction of $2.3 \pm 0.6$, i.e., that likely each of the sources has (at least) two stellar companions.
\citet{2019MNRAS.488.2480R} investigated the potential for orbit hardening through the EKL-mechanism. Their findings indicate that, starting with a cutoff period of $\sim9$\,months, as proposed by \citet{2017A&A...599L...9S}, the EKL-mechanism falls short in replicating the population of short-period binaries as observed by \citet{2012Sci...337..444S} for $\sim2-5$\,Myr old clusters. With our new observations we discard distributions with cutoff periods $\lesssim$\,\pcutoffMoSig\,days with 68\% confidence. New simulations are needed to test whether the EKL-mechanism can build up a sufficient population of short period binaries within $\sim$10$^{6}$\,yr when the initial period is $\sim3$\,months. 

\emph{Dynamical Interactions:} 
\citet{2021A&A...645L..10R} state that due to the low stellar densities observed in the studied clusters, stellar encounters are expected to be infrequent. Consequently, mechanisms like binary-binary or single-binary interactions \citep[][]{2002MNRAS.336..705B, 2009MNRAS.392..590B} were not expected to play a substantial role in the formation of close binaries. 
Nevertheless, the evidence for many runaway stars in M17 \citep[][]{2024A&A...681A..21S} and the fact that we expect several wide binaries in this cluster could indicate that these interactions are more important than thought before. 
If that was the case, dynamical interactions would be a hardening mechanism, resulting in runaways \citep[e.g. disrupted binaries;][]{2012MNRAS.424..272P} and close binaries.

\emph{Tidal circularization and orbital decay:} \citet{2015ApJ...810...61M} report that the highly eccentric young binaries in their sample nearly fill their Roche lobes at periastron. These young sources tidally circularize and decay toward shorter periods on timescales of a few Myr. These two effects together could explain how young stars that are born in wide orbits harden in the first million years of evolution to represent the properties observed in $\sim2-5$\,Myr clusters. In order to test this scenario, observations of additional epochs are needed in order to find the orbital solutions of the systems and to test if there is a trend of eccentricity with cluster age.

Models of massive-star populations including binary stars predict that only a small percentage of the binaries will interact within the first 3\,Myrs of evolution. In most cases where interaction does occur, the stars will merge to produce a blue straggler \citep[][]{2015ApJ...805...20S}. 
Our findings about the hardening of binaries during the initial million years, unless specifically concerned with early blue-straggler formation, are not expected to influence predictions made by binary-population synthesis models. Therefore, starting with the initial conditions described in \citet{2012Sci...337..444S}, remains a valid approach if one is interested in the outcome of binary evolution after the initial Myrs.

\section{Conclusions}\label{P5:conclusions}

With the goal of testing the early main-sequence migration scenario for the formation of massive binary stars, we performed a radial-velocity analysis of a population of 21 young stars in the $\sim0.7$\,Myr old giant \Hii\ region M17. 
% 15 stars with masses ranging from 3 to 30\,\msun have multi-epoch spectroscopy with 2 to 5 observations and for 20 stars we were able to measure RVs in at least one epoch. 
Our results can be summarized as follows: 

   \begin{itemize}
      \item We find that both single epoch and multi-epoch analyses with typical sample sizes of around 10 stars lead to the conclusion that there is a lack of short period binaries in the young region M17 relative to $\sim2-5$\,Myr old clusters. 

      \item The derived radial velocity dispersion for the extended sample presented in this paper is \srv\,$=$\,\srvM\,\kms. This is  higher than the \srv\,$=5.6\pm0.2$\,\kms\ derived for 11 stars by \citet{2017A&A...599L...9S}. The difference is mainly due to the inclusion of three single-line spectroscopic binaries (SB1; B150, B253, and B86) in the new sample. However, the new somewhat higher value observed for the young stars in M17 remains strikingly low in comparison with the $\sim30$\,\kms\ observed in older clusters with similar mass in the Milky Way. 

      \item The \srv\ obtained for samples containing stars with and without disks agree within the errors. This implies that the low \srv\ reported by \citet{2017A&A...599L...9S} is not due to the initial sample selection that aimed at targeting stars with IR-excess \citep{2017A&A...604A..78R}.

      \item We derive a spectroscopic observed binary fraction for M17 of \fbin\,$=$\,27\%. 
      Correcting for observational biases, we estimate an intrinsic binary fraction of \intrFbin\%, consistent with the universal value \citep[][]{2012Sci...337..444S}. 
      
      \item Assuming \fbin\,$=$\,70\%, the most likely \pcutoff\ to explain the observed \srv\ in M17 is \pcutoffM\,days. We reject distributions with \pcutoff\,$<$\,\pcutoffMoSig (\pcutoffMtSig) days with 68(95)\% confidence, respectively. 
      The derived value of \fbin\,$=$\,\intrFbin\% leads to a best-fit distribution with \pcutoff\,$=$\,\pcutoffMfeighty\,days and distributions with \pcutoff\,$<$\,\pcutoffMfeightyOsig (\pcutoffMfeightyTsig) days can be rejected with 68(95)\% confidence.
            
      \item The transverse velocity dispersion of stellar systems in star clusters as derived from their proper motions ($\sigma_{\rm 1D, RA}$, $\sigma_{\rm 1D, DEC}$) is not affected by binary motions and only reflects random motions. We find that it does not vary with age and ranges between 0.4 and 2.3\,\kms. We stress the importance of using the tangential velocity dispersion, rather than relying on the radial velocity dispersion dominated by binary motions, when estimating the masses of star clusters thought to be in virial equilibrium. Using the latter would lead to an overestimate of the star cluster mass.

   \end{itemize}

We confirm that 3 to 30 \msun\ stars in the young massive cluster M17 fit in a trend of binary hardening on the pre- and early main-sequence first presented by \citet{2021A&A...645L..10R}. 
This early migration process shows that early binary evolution together with possibly dynamical interactions are key processes in the formation of gravitational wave sources. 
Though our findings are statistically significant, the validity of the trend deserves further scrutiny. Larger and unbiased samples of young massive stars are needed to address the dependence of the underlying distributions on primary mass \citep[see][]{2017ApJS..230...15M}. Additional epoch observations will allow us to find the exact periods, mass ranges and eccentricities of young systems like M17, to be able to fully constrain close binary formation theories. The more precise the trend of hardening with time is established, the better the trend itself can be used to age-date young clusters from relatively simple spectroscopic data sets.

\begin{acknowledgements}

We would like to thank Max Moe for his valuable suggestions which have improved the quality of this manuscript.
We thank Tim Michael Heinz Wolf for the assistance developing \texttt{RVFitter}, the code used to measure radial velocities, and Stefano Rinaldi and Morgan Fouesneau for valuable discussions about the statistical approach of this paper. 
MCRT acknowledges support by the German Aerospace Center (DLR) and the Federal Ministry for Economic Affairs and Energy (BMWi) through program 50OR2314 ‘Physics and Chemistry of Planet-forming disks in extreme environments’.
The research leading to these results has received funding from the European Research Council (ERC) under the European Union's Horizon 2020 research and innovation programme (grant agreement numbers 772225: MULTIPLES).
This work is based on observations collected at the European Organization for Astronomical Research in the Southern Hemisphere under ESO programs 60.A-9404(A), 085.D-0741, 089.C-0874(A), 091.C-0934(B) and 103.D-0099. 
This research has made use of the SIMBAD database, operated at CDS, Strasbourg, France. This research made use of Astropy \citep{astropy:2013, astropy:2018, astropy:2022}, NumPy \citep{2020Natur.585..357H}, and Matplotlib \citep{2007CSE.....9...90H}.

\end{acknowledgements}

%-------------------------------------------------------------------
\bibliographystyle{aa}
\bibliography{references}

%-------------------------------------------------------------------

\begin{appendix}

\section{Log of Observations}
\label{P5:app:obslog}

\renewcommand{\arraystretch}{1.5}
\setlength{\tabcolsep}{14pt}
\onecolumn
\begin{longtable}{cccccc}
\caption{Log of the observations used in this paper. The first column lists the name of the object, the second and third list the right ascension and declination (J2000), the fourth and fifth list the date of observation and it's corresponding julian date and the sixth column lists the barycentric radial velocity for each observation.}
% \begin{tabular}
\\ 
\hline
\hline
Object & RA & DEC & Date & HJD & Barycor \\
 & (J2000) & (J2000) & YYY-MM-DD & -2\,400\,000.5\,days & km/s \\
\hline
326 & 275.108 & -16.143 & 2019-07-06 & 58670.042189662 & -4.154 \\
326 & 275.107 & -16.143 & 2019-07-09 & 58673.019478376 & -5.573 \\
326 & 275.108 & -16.143 & 2019-08-09 & 58704.003683766 & -19.223 \\
B111 & 275.143 & -16.171 & 2013-07-17 & 56490.115837414 & -9.855 \\
B111 & 275.144 & -16.17 & 2019-04-11 & 58584.399746601 & 28.46 \\
B111 & 275.145 & -16.17 & 2019-04-13 & 58586.365560726 & 28.233 \\
B111 & 275.144 & -16.17 & 2019-06-05 & 58639.242172575 & 10.533 \\
B150 & 275.132 & -16.195 & 2019-07-06 & 58670.188642478 & -4.554 \\
B150 & 275.132 & -16.195 & 2019-07-09 & 58673.050560242 & -5.632 \\
B150 & 275.133 & -16.195 & 2019-08-09 & 58704.032736219 & -19.297 \\
B164 & 275.128 & -16.169 & 2013-07-17 & 56490.123330743 & -9.885 \\
B164 & 275.129 & -16.169 & 2019-06-05 & 58639.266762098 & 10.451 \\
B164 & 275.129 & -16.169 & 2019-06-07 & 58641.289724467 & 9.433 \\
B164 & 275.129 & -16.169 & 2019-07-10 & 58674.021019356 & -6.051 \\
B181 & 275.127 & -16.177 & 2019-08-04 & 58699.180929747 & -17.792 \\
B205 & 275.121 & -16.186 & 2019-06-05 & 58639.249821618 & 10.499 \\
B205 & 275.121 & -16.186 & 2019-06-06 & 58640.216834524 & 10.124 \\
B205 & 275.121 & -16.186 & 2019-07-06 & 58670.222459632 & -4.661 \\
B213 & 275.119 & -16.157 & 2019-08-04 & 58699.221264151 & -17.886 \\
B215 & 275.119 & -16.204 & 2019-07-30 & 58694.214764089 & -15.8 \\
B215 & 275.119 & -16.204 & 2019-08-04 & 58699.136502983 & -17.678 \\
B215 & 275.119 & -16.204 & 2019-09-27 & 58753.99260901 & -29.645 \\
B228 & 275.116 & -16.167 & 2019-07-06 & 58670.029573872 & -4.123 \\
B228 & 275.116 & -16.166 & 2019-07-08 & 58672.989984377 & -5.514 \\
B228 & 275.116 & -16.167 & 2019-08-03 & 58698.119743304 & -17.197 \\
B230 & 275.116 & -16.184 & 2019-07-30 & 58694.173671121 & -15.697 \\
B230 & 275.116 & -16.184 & 2019-08-01 & 58696.095419848 & -16.321 \\
B230 & 275.116 & -16.184 & 2019-08-29 & 58724.021699753 & -25.671 \\
B234 & 275.114 & -16.17 & 2019-08-05 & 58700.098835423 & -17.97 \\
B243 & 275.11 & -16.168 & 2019-07-30 & 58694.131399116 & -15.579 \\
B243 & 275.11 & -16.168 & 2019-08-03 & 58698.185489372 & -17.404 \\
B243 & 275.11 & -16.168 & 2019-09-25 & 58751.063829271 & -29.727 \\
B243 & 275.111 & -16.167 & 2012-07-06 & 56114.188166478 & -4.877 \\
B243 & 275.111 & -16.168 & 2013-07-17 & 56490.140132366 & -9.944 \\
B253 & 275.108 & -16.185 & 2019-06-05 & 58639.284850709 & 10.387 \\
B253 & 275.108 & -16.185 & 2019-06-07 & 58641.272525638 & 9.475 \\
B253 & 275.108 & -16.185 & 2019-07-09 & 58673.114315431 & -5.816 \\
B268 & 275.106 & -16.172 & 2012-07-06 & 56114.236476209 & -5.022 \\
B268 & 275.106 & -16.172 & 2012-07-06 & 56114.259057509 & -5.084 \\
B268 & 275.106 & -16.172 & 2013-07-17 & 56490.169532141 & -10.034 \\
B268 & 275.105 & -16.173 & 2019-05-30 & 58633.335883405 & 12.986 \\
B268 & 275.105 & -16.173 & 2019-05-31 & 58634.306223739 & 12.617 \\
B268 & 275.105 & -16.172 & 2019-07-29 & 58693.173052216 & -15.269 \\
B269 & 275.107 & -16.187 & 2019-06-14 & 58648.248693726 & 6.167 \\
B269 & 275.106 & -16.187 & 2019-07-12 & 58676.100162493 & -7.229 \\
B272 & 275.105 & -16.162 & 2019-08-05 & 58700.057678609 & -17.852 \\
B275 & 275.105 & -16.174 & 2019-06-05 & 58639.314933648 & 10.285 \\
B275 & 275.105 & -16.174 & 2019-06-06 & 58640.24300323 & 10.024 \\
B275 & 275.105 & -16.174 & 2019-07-09 & 58673.091584056 & -5.765 \\
B289 & 275.102 & -16.146 & 2010-09-17 & 55456.02383212 & -29.069 \\
B289 & 275.102 & -16.145 & 2012-07-06 & 56114.290051974 & -5.163 \\
B289 & 275.101 & -16.145 & 2019-05-18 & 58621.384886585 & 17.949 \\
B289 & 275.101 & -16.145 & 2019-06-24 & 58658.203572001 & 1.356 \\
B290 & 275.101 & -16.152 & 2019-07-09 & 58673.036654984 & -5.614 \\
B290 & 275.101 & -16.153 & 2019-07-06 & 58670.057762187 & -4.194 \\
B290 & 275.101 & -16.153 & 2019-08-09 & 58704.018851714 & -19.267 \\
B293 & 275.1 & -16.138 & 2019-07-06 & 58670.081535306 & -4.254 \\
B293 & 275.1 & -16.138 & 2019-07-09 & 58673.084651741 & -5.735 \\
B311 & 275.094 & -16.143 & 2013-07-17 & 56490.096661869 & -9.82 \\
B311 & 275.095 & -16.143 & 2019-05-01 & 58604.365915942 & 24.02 \\
B311 & 275.095 & -16.143 & 2019-05-02 & 58605.39364885 & 23.639 \\
B311 & 275.096 & -16.143 & 2019-06-05 & 58639.302147351 & 10.33 \\
B331 & 275.091 & -16.188 & 2012-07-07 & 56115.276048548 & -5.623 \\
B336 & 275.088 & -16.194 & 2019-07-06 & 58670.205953578 & -4.628 \\
B336 & 275.089 & -16.194 & 2019-07-09 & 58673.067513681 & -5.696 \\
B336 & 275.088 & -16.194 & 2019-08-09 & 58704.052589863 & -19.37 \\
B337 & 275.089 & -16.178 & 2013-07-16 & 56489.184642895 & -9.617 \\
B337 & 275.089 & -16.179 & 2019-05-01 & 58604.323445446 & 24.115 \\
B337 & 275.088 & -16.178 & 2019-08-03 & 58698.078349757 & -17.112 \\
B86 & 275.151 & -16.181 & 2019-07-06 & 58670.069755122 & -4.2 \\
B86 & 275.151 & -16.181 & 2019-07-09 & 58673.005433982 & -5.526 \\
B86 & 275.145 & -16.185 & 2019-08-03 & 58698.137451515 & -17.237 \\
B93 & 275.148 & -16.183 & 2019-08-08 & 58703.990430332 & -19.176 \\
B93 & 275.148 & -16.183 & 2019-06-24 & 58658.266146998 & 1.21 \\
CEN55 & 275.121 & -16.183 & 2019-08-05 & 58700.139589769 & -18.085 \\
\hline
% \end{tabular}
\end{longtable}
\twocolumn

\section{Spectral classification}\label{P5:app:spectral classification}
\subsection{OB stars}

\subsection*{B86 - B9\,III}
This star has relatively strong Mg\,{\sc ii} absorption at 4481\,\AA\ compared to the neighboring \hea\ at 4471\,\AA\ line, which indicates that B86 is a B8-B9 star. Since the spectrum shows Fe\,{\sc ii} 4233\,\AA\ and considering the ratio of Mg\,{\sc ii} to the bordering \hea\ line, the star is classified as B9. The luminosity class is III since the Si\,{\sc ii} lines at 4128-30\,\AA\ are of similar strength compared to \hea\ 4026\,\AA.

\subsection*{B93 - B2\,V}
B93 shows no \heb\ lines, but only \hea\ lines in the spectrum, where \hea\,4144\,\AA\ is stronger than \hea\,4121\,\AA, indicating early B, but not as hot as B0 or B1. The Si\,{\sc iii} 4552\,\AA\ ratio with Si\,{\sc ii} 4128\,\AA\ and with Mg\,{\sc ii} 4481\,\AA\ are similar to that of spectral type B2. This is confirmed by the presence of O\,{\sc ii} 4070-76\,\AA. Since N\,{\sc ii} 3995\,\AA\ is not detected, the luminosity class is V. 

\subsection*{B181 - B0\,V}
We detect one strong \heb\ line in this star at 4686\,\AA\ and some \hea\ lines. This corresponds to spectral type B0, as we do not detect \heb\ 4541\,\AA\ as is expected for late O-type stars. The \heb\ line is stronger than \hea\ 4711\,\AA, indicating luminosity class V. 

\subsection*{B205 - B2\,V}
The stacked spectrum of B205 has strong \hea\ lines and no \heb\ lines, which indicates that the star must be an early B type star but cooler than B0. Based on the ratio of \hea\ 4142\,\AA\ and \hea\ 4144\,\AA\ (which is 1:2) and the detection of Si\,{\sc iii} at 4552, 4568 and 4575\,\AA, the star is classified as B2. Since the hydrogen lines are relatively broad and O\,{\sc ii} is not detected at 4350\,\AA, the luminosity class is V. 

\subsection*{B213 - B3\,III-V}
The spectrum shows strong \hea\ lines compared to H\,{\sc i}, but no \heb\ lines. The Mg\,{\sc ii} at 4481\,\AA\ is weak compared to \hea\ at 4471\,\AA\ and Si\,{\sc iv}\,4089\,\AA\ is not detected, suggesting a B3 classification. The spectrum shows relatively broad H\,{\sc i} and \hea\ lines, but is too noisy to distinguish between luminosity class III and V. 

\subsection*{B234 - B3\,III-V}
The spectrum of B234 is similar to B213 and we arrive at the same spectral classification. 

\subsection*{B272 - B7\,III}
B272 is of late-B type since the spectrum shows a relatively strong Mg\,{\sc ii} feature at 4481\,\AA. However, \hea\ at 4471\,\AA\ is stronger than Mg\,{\sc ii} in the ratio 3:2. This indicates a spectral type B7. The luminosity class III is based on the similar strength of the Si\,{\sc ii} lines at 4128-30\,\AA\ and \hea\ 4026\,\AA.

\subsection*{CEN55 - B2\,III}
The spectrum shows similar characteristics as B93 and B205 that have a spectral type B2. CEN55 shows N\,{\sc ii} 3995\,\AA\ in the spectrum corresponding to luminosity class III.

\subsection*{326 - O6\,III + B1\,V}
Visual inspection of the spectrum of 326 reveals that it is a double-lined spectroscopic binary (SB2). The last epoch shows a relatively large spread between the absorption lines in a few \hea\ lines and \ha. This epoch is used to give a tentative indication of spectral type. 

One of the stars shows strong \heb\ lines and exhibits characteristics of an O6 star; \heb\ at 4200\,\AA\ is of similar strength as \hea\ at 4026\,\AA\ and N\,{\sc iii} 4634-40-42\,\AA\ is in emission. The luminosity class is III inferred from a similar strength of \heb\ 4541\,\AA\ and \hea\ 4471\,\AA. 

The second star does not seem to have \heb\ lines. Its Mg\,{\sc ii} 4481\,\AA\ is relatively weak compared to \hea\ 4471\,\AA\, where the latter is also weaker than in the O6\,III star. \hea\ 4144\,\AA\ is similarly strong as \hea\ 4121, which is stronger than Si\,{\sc ii} 4128-30\,\AA, therefore this star is an early B type. The spectrum shows O\,{\sc ii} 4070-76\,\AA, Si\,{\sc iii} at 4552, 4568 and 4575\,\AA\ and Si\,{\sc iv}\,4089\,\AA, which is found for B1\,V stars. 

\subsection{Cooler stars}
The cooler stars in the sample were given a general classification using distinct features in \cite{gray2000digital}. Strong hydrogen lines indicate an A-type star, where for F-type stars the Ca\,{\sc ii} K-line grows stronger and the G-band starts to appear. The G-band is strongest in the G-type stars in the sample. 

\begin{figure*}
\centering
\includegraphics[width=0.9\hsize]{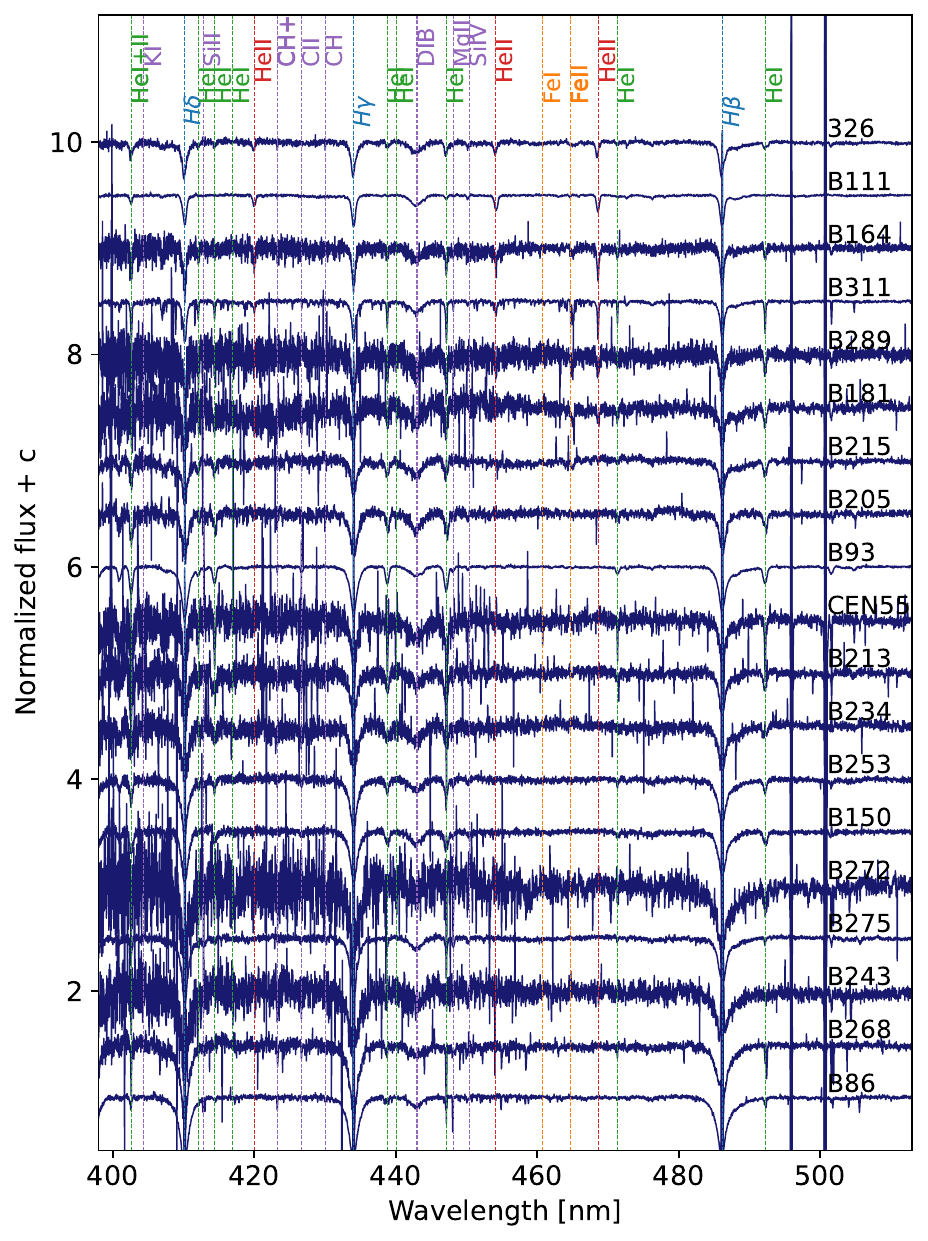}
  \caption{Spectra of the stars at the wavelengths used for classification. The stars are listed from hottest (top) to coolest (bottom). B337 and B331 are not included due the high extinction in this wavelength range. }
     \label{P5:fig:spectra}
\end{figure*}

\section{RV measurements and lines used per star}
\onecolumn
\begin{longtable}{ccccccccc}
\caption{Radial-velocity measurements for our sources per epoch. Column 2 shows the Julian Date the ISM velocity measured from interstellar absorption lines is listed in column 3 and the radial velocity of the star in the last column.}\\

\hline
\hline
Star & HJD & RV$_{\rm ISM}$ & RV \\
 & -2\,400\,000.5\,days & \kms & \kms \\
\hline
326 & 58670.042189662 & -3.0 $\pm$ 0.2 & --- \\
 & 58673.019478376 & -4.7 $\pm$ 0.2 & --- \\
 & 58704.003683766 & -3.7 $\pm$ 0.2 & --- \\
B111 & 56490.115837414 & -3.0 $\pm$ 0.1 & 22 $\pm$ 1 \\
 & 58584.399746601 & -4.3 $\pm$ 0.1 & 14 $\pm$ 1 \\
 & 58586.365560726 & -4.3 $\pm$ 0.1 & 20 $\pm$ 1 \\
 & 58639.242172575 & -4.6 $\pm$ 0.1 & 14 $\pm$ 1 \\
B150 & 58670.188642478 & -3.1 $\pm$ 0.2 & 42 $\pm$ 5 \\
 & 58673.050560242 & -6.1 $\pm$ 0.2 & 55 $\pm$ 4 \\
 & 58704.032736219 & -3.8 $\pm$ 0.2 & 16 $\pm$ 4 \\
B164 & 56490.123330743 & -2.6 $\pm$ 0.2 & 11 $\pm$ 1 \\
 & 58639.266762098 & -2.3 $\pm$ 0.2 & 14 $\pm$ 1 \\
 & 58641.289724467 & -3.7 $\pm$ 0.2 & 14 $\pm$ 1 \\
 & 58674.021019356 & -2.9 $\pm$ 0.2 & 15 $\pm$ 1 \\
B181 & 58699.180929747 & -1.3 $\pm$ 0.3 & -10 $\pm$ 3 \\
B205 & 58639.249821618 & -1.8 $\pm$ 0.3 & 4 $\pm$ 2 \\
 & 58640.216834524 & 1.0 $\pm$ 0.3 & 23 $\pm$ 2 \\
 & 58670.222459632 & -1.5 $\pm$ 0.3 & 18 $\pm$ 2 \\
B213 & 58699.221264151 & -5.1 $\pm$ 0.2 & 0 $\pm$ 5 \\
B215 & 58694.214764089 & -1.0 $\pm$ 0.2 & 2 $\pm$ 3 \\
 & 58699.136502983 & -1.0 $\pm$ 0.2 & 6 $\pm$ 2 \\
 & 58753.99260901 & -2.6 $\pm$ 0.2 & 1 $\pm$ 2 \\
B228 & 58670.029573872 & -26.1 $\pm$ 0.4 & -12 $\pm$ 2 \\
 & 58672.989984377 & -23.9 $\pm$ 0.4 & -14 $\pm$ 2 \\
 & 58698.119743304 & -23.9 $\pm$ 0.4 & -14 $\pm$ 2 \\
B230 & 58694.173671121 & -2.6 $\pm$ 0.5 & 6 $\pm$ 1 \\
 & 58696.095419848 & 0.6 $\pm$ 0.5 & 8 $\pm$ 1 \\
 & 58724.021699753 & -2.8 $\pm$ 0.5 & 6 $\pm$ 1 \\
B234 & 58700.098835423 & -1.1 $\pm$ 0.4 & -4 $\pm$ 4 \\
B243 & 56114.188166478 & 1.6 $\pm$ 0.6 & 9 $\pm$ 6 \\
 & 56490.140132366 & 0.2 $\pm$ 0.6 & 0 $\pm$ 6 \\
 & 58694.131399116 & 1.6 $\pm$ 0.6 & -4 $\pm$ 6 \\
 & 58698.185489372 & 0.0 $\pm$ 0.6 & 5 $\pm$ 6 \\
 & 58751.063829271 & -1.2 $\pm$ 0.6 & -7 $\pm$ 6 \\
B253 & 58639.284850709 & 2.7 $\pm$ 0.3 & 1 $\pm$ 4 \\
 & 58641.272525638 & 2.0 $\pm$ 0.3 & 6 $\pm$ 4 \\
 & 58673.114315431 & 0.2 $\pm$ 0.3 & 22 $\pm$ 4 \\
B268 & 56114.259057509 & 4.8 $\pm$ 0.3 & 9 $\pm$ 3 \\
 & 56490.169532141 & 4.4 $\pm$ 0.3 & 8 $\pm$ 3 \\
 & 58633.335883405 & -0.7 $\pm$ 0.3 & 1 $\pm$ 3 \\
 & 58634.306223739 & -1.0 $\pm$ 0.3 & 0 $\pm$ 3 \\
 & 58693.173052216 & 7.8 $\pm$ 0.3 & -5 $\pm$ 3 \\
B269 & 58648.248693726 & 0.1 $\pm$ 0.4 & 13 $\pm$ 2 \\
 & 58676.100162493 & -0.7 $\pm$ 0.4 & 13 $\pm$ 2 \\
B272 & 58700.057678609 & -2.5 $\pm$ 0.4 & 3 $\pm$ 0 \\
B275 & 58639.314933648 & -1.2 $\pm$ 0.2 & 17 $\pm$ 2 \\
 & 58640.24300323 & -2.2 $\pm$ 0.2 & 12 $\pm$ 2 \\
 & 58673.091584056 & -1.8 $\pm$ 0.2 & 5 $\pm$ 2 \\
B289 & 55456.02383212 & -0.5 $\pm$ 0.5 & 7 $\pm$ 2 \\
 & 56114.290051974 & -1.7 $\pm$ 0.5 & 12 $\pm$ 2 \\
 & 58621.384886585 & -0.8 $\pm$ 0.5 & 10 $\pm$ 2 \\
 & 58658.203572001 & 0.1 $\pm$ 0.5 & 9 $\pm$ 1 \\
B290 & 58670.057762187 & -5.4 $\pm$ 0.2 & 11 $\pm$ 2 \\
 & 58673.036654984 & -3.7 $\pm$ 0.2 & 1 $\pm$ 2 \\
 & 58704.018851714 & -2.9 $\pm$ 0.2 & -4 $\pm$ 2 \\
B293 & 58670.081535306 & -38.0 $\pm$ 1.0 & -160 $\pm$ 4 \\
 & 58673.084651741 & -38.0 $\pm$ 1.0 & -156 $\pm$ 4 \\
B311 & 56490.096661869 & -0.5 $\pm$ 0.2 & 8 $\pm$ 1 \\
 & 58604.365915942 & -2.6 $\pm$ 0.2 & 4 $\pm$ 1 \\
 & 58605.39364885 & -3.4 $\pm$ 0.2 & 0 $\pm$ 1 \\
 & 58639.302147351 & -2.3 $\pm$ 0.2 & 3 $\pm$ 1 \\
% B331 & 58670.205953578 & 14 $\pm$ 2\footnote{measurement from \cite{2017A&A...599L...9S}} & 2.1 $\pm$ 0.3 \footnote{average of all ISM RV measurements}\\
B331 & 56115.276048548 & 10.2 $\pm$ 0.3 & 24 $\pm$ 3 \\
B336 & 58670.205953578 & -26.7 $\pm$ 0.5 & 27 $\pm$ 3 \\
 & 58673.067513681 & -16.9 $\pm$ 0.5 & 47 $\pm$ 3 \\
 & 58704.052589863 & -18.7 $\pm$ 0.5 & 38 $\pm$ 4 \\
B337 & 56489.184642895 & 4.0 $\pm$ 1.0 & 5 $\pm$ 3 \\
 & 58604.323445446 & -5.0 $\pm$ 1.0 & -8 $\pm$ 3 \\
 & 58698.078349757 & -2.0 $\pm$ 1.0 & -5 $\pm$ 3 \\
B86 & 58670.069755122 & -0.6 $\pm$ 0.3 & -8 $\pm$ 3 \\
 & 58673.005433982 & -3.4 $\pm$ 0.3 & -41 $\pm$ 3 \\
 & 58698.137451515 & -2.7 $\pm$ 0.3 & -35 $\pm$ 3 \\
B93 & 58658.266146998 & -0.6 $\pm$ 0.1 & 4 $\pm$ 1 \\
 & 58703.990430332 & -0.1 $\pm$ 0.1 & 4 $\pm$ 1 \\
CEN55 & 58700.139589769 & -3.3 $\pm$ 0.5 & 18 $\pm$ 2 \\
\hline
\hline
\label{P5:tab:RV_measurements}
\end{longtable}

\section{Lines used to measure RV per star}\label{P5:app:lines_used}
\renewcommand{\arraystretch}{1.1}
\setlength{\tabcolsep}{1pt}
\begin{sidewaystable*}
\caption{Spectral lines used per star to measure the radial velocity. The central wavelength shown in this table is the exact central value used to calculate the radial velocity.}
\begin{tabular}{l|c|c|c|c|c|c|c|c|c|c|c|c|c|c|c|c|c|c|c|c|c|c|c|c|c|c|}
\toprule
 & B111 & B150 & B164 & B181 & B205 & B213 & B215 & B228 & B230 & B234 & B243 & B253 & B268 & B269 & B272 & B275 & B289 & B290 & B293 & B311 & B331 & B336 & B337 & B86 & B93 & CEN55 \\
\midrule
CII 4267.26 &   &   &   &   &   &   &   &   &   &   &   &   &   &   &   &   &   &   &   & x &   &   &   &   & x &   \\
MgII 4481.13 &   &   &   &   & x &   &   &   &   &   &   &   &   &   &   &   &   & x &   & x &   &   &   & x &   &   \\
SiII 4128.0 &   &   &   &   &   &   &   &   &   &   &   &   &   &   &   &   &   &   &   &   &   &   &   & x &   &   \\
SiIII 4552.62 &   &   &   &   &   &   &   &   &   &   &   &   &   &   &   &   &   &   &   & x &   &   &   &   &   &   \\
OII 4661.6332 &   &   &   &   &   &   &   &   &   &   &   &   &   &   &   &   &   &   &   & x &   &   &   &   &   &   \\
Ba-12 3750.151 &   &   &   &   &   &   &   &   &   &   &   &   &   &   &   &   &   & x &   &   &   &   &   &   & x &   \\
Ba-11 3770.633 &   &   &   &   &   &   &   &   &   &   &   &   &   &   &   &   &   & x &   &   &   &   &   &   & x &   \\
Ba-10 3797.909 &   &   &   &   & x &   &   &   &   &   &   &   &   &   &   &   &   &   &   &   &   &   &   &   & x &   \\
Ba-9 3835.397 &   &   &   &   & x &   &   &   &   & x &   &   &   &   &   &   &   &   &   &   &   &   &   &   &   &   \\
$H\eta$ 3889.064 &   &   &   &   &   &   &   &   &   & x &   &   &   &   &   &   &   &   &   &   &   &   &   &   &   &   \\
$H\epsilon$ 3970.075 &   &   &   & x &   &   &   &   &   &   & x &   &   &   &   &   &   &   & x &   &   &   &   &   &   &   \\
$H\delta$ 4101.734 &   &   &   &   & x & x &   &   &   & x & x &   &   &   &   &   &   &   &   &   &   &   &   &   &   &   \\
$H\gamma$ 4340.472 &   &   &   & x & x & x &   &   &   & x & x &   & x &   &   &   &   &   &   &   &   &   &   &   & x &   \\
$H\beta$ 4861.35 &   &   &   & x & x & x &   &   &   &   &   &   & x &   &   &   &   &   &   &   &   &   &   &   &   &   \\
H$\alpha$ 6562.79 &   &   &   & x &   &   &   &   &   &   &   &   &   &   &   &   &   &   &   &   &   &   &   &   &   &   \\
Pa-11 8862.89 &   &   &   & x & x & x &   &   & x & x & x &   & x & x & x &   &   &   &   &   & x &   & x &   &   &   \\
Pa-10 9015.3 &   &   &   &   &   &   &   &   &   &   &   &   & x &   & x &   &   &   & x &   &   &   &   &   &   &   \\
Pa-9 9229.7 &   &   &   & x & x & x &   &   &   & x &   &   & x &   &   &   &   &   &   &   &   &   & x &   &   &   \\
Pa-7 10049.8 &   &   &   & x & x & x &   &   & x & x & x &   &   & x & x &   &   &   & x &   &   &   & x &   &   &   \\
HeI 3819.6074 &   &   &   &   &   &   &   &   &   &   &   &   &   &   &   &   &   &   &   &   &   & x &   &   & x &   \\
HeI 3888.648 & x & x &   &   &   & x &   &   &   & x &   & x &   &   &   &   &   & x &   &   &   &   &   &   &   &   \\
HeI+II 4026.1914 & x & x &   &   & x &   &   &   &   & x &   &   &   &   &   &   &   &   &   &   &   &   &   &   & x &   \\
HeI 4387.9296 &   & x &   & x & x &   &   &   &   &   &   & x &   &   &   &   &   &   &   &   &   &   &   &   & x & x \\
HeI 4471.4802 & x &   & x &   & x & x &   &   &   & x &   & x &   &   & x &   &   &   &   &   &   &   &   &   & x & x \\
HeI 4921.9313 &   & x &   & x & x &   &   &   &   & x &   & x &   &   &   &   &   &   &   &   &   &   &   &   & x & x \\
HeI 5015.6783 &   &   & x & x & x &   &   &   &   &   &   &   &   &   & x &   &   &   &   &   &   &   &   &   &   &   \\
HeI 5875.621 & x &   & x & x & x & x & x &   &   &   &   & x &   &   &   & x & x &   &   &   &   &   &   &   &   & x \\
HeI 6678.151 &   &   & x & x & x & x & x &   & x & x &   & x &   &   &   & x & x &   &   &   &   &   &   &   & x &   \\
HeI 7065.19 &   &   & x & x & x &   & x &   &   &   &   &   &   &   &   &   & x &   &   &   &   &   &   &   &   & x \\
HeII 4199.6 & x &   &   &   &   &   &   &   &   &   &   &   &   &   &   &   &   &   &   &   &   &   &   &   &   &   \\
HeII 4541.0 &   &   &   &   &   &   &   &   &   &   &   &   &   &   &   &   &   & x &   &   &   &   &   &   &   &   \\
HeII 4685.7 & x &   & x &   &   &   &   &   &   &   &   &   &   &   &   &   &   &   &   &   &   &   &   &   &   &   \\
\bottomrule
\end{tabular}

\end{sidewaystable*}

\newpage

\section{Eccentricity}\label{P5:app:eccentricity}

\begin{figure}[h]
    \centering
    \begin{subfigure}%[b]{0.48\linewidth}        %% or \columnwidth
        \centering
        \includegraphics[width=0.48\linewidth]{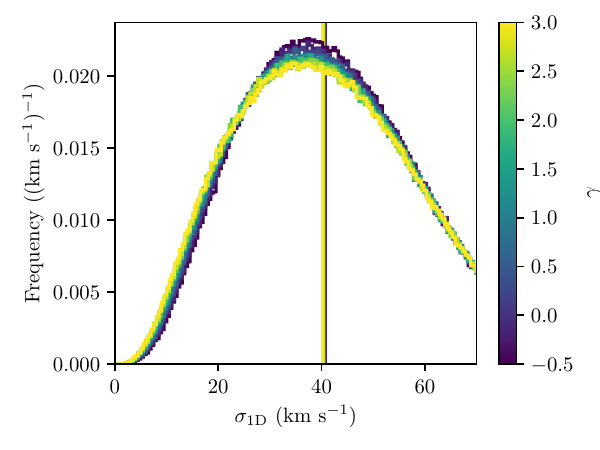}
    \end{subfigure}
    \begin{subfigure}%[b]{0.48\linewidth}        %% or \columnwidth
        \centering
        \includegraphics[width=0.48\linewidth]{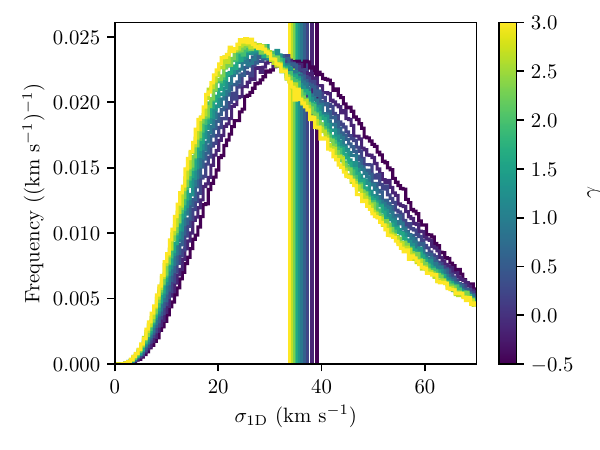}
        \label{P5:fig:B}
    \end{subfigure}
    \caption{Effect of changing the power index of the eccentricity, $\gamma$, from -0.5 to 3. \textit{Left:} Keeping the dependency of the eccentricity distribution on the period of the binary. For $P<4$ days circular orbits are assumed, for periods between 4 and 6 days, the eccentricities are sampled from $\text{pdf}(e) \propto e^{-0.5}$, with $0 \leq e < 0.5$; for periods longer than 6 days the same distribution is used, but with $0 \leq e < 0.9$. \textit{Right:} Allowing all orbits to be eccentric.}
    \label{P5:fig:eccentricity}
\end{figure}

In this section we focus in the effect of varying the eccentricity $e$ on the observed \srv\ of clusters. We follow the approach described in \citet{2017A&A...599L...9S}. 
For the basic properties of the population we adopt the values of \citet{2012Sci...337..444S} for Galactic young clusters. The eccentricity distribution depends on the period of the binary. For $P<4$ days circular orbits are assumed, for periods between 4 and 6 days, the eccentricities are sampled from $\text{pdf}(e) \propto e^{-0.5}$, with $0 \leq e < 0.5$; for periods longer than 6 days the same distribution is used, but with $0 \leq e < 0.9$.
The left panel of Figure~\ref{P5:fig:eccentricity} shows the effect of varying the power index of the eccentricity, $\gamma$, from -0.5 (corresponding to \citet{2012Sci...337..444S} empirical value) to 3 (strong preference for eccentric orbits) in steps of 0.35 on the \srv\ of different populations. The \srv\ value varies from $\sim41$\,\kms\ to $\sim40$\,\kms. In the right panel we show the effect of changing the eccentricity power as below, but removing the eccentricity dependence on the binary period. In this extreme case the eccentricity varies from $\sim40$\,\kms\ to $\sim34$\,\kms, in none of the cases the effect of eccentricity is enough to account for the low \srv\ observed in M17.

\end{appendix}

\end{document}